# Retinex-qDPC: automatic background rectified quantitative differential phase contrast imaging

Shuhe Zhang,[1, 2, **], Tao Peng,[1] Zeyu Ke,[1] Han Yang,[1] Tos T. J. M. Berendschot,[2] and Jinhua Zhou,[1,3, *]

1. *School of Biomedical Engineering, Anhui Medical University, Hefei 230032, China.*
2. *University Eye Clinic Maastricht, Maastricht University Medical Center + , P.O. Box 5800, Maastricht, AZ 6202, the Netherlands*
3. *Anhui Provincial Institute of Translational Medicine, Anhui Medical University, Hefei 230032, China.*

*<u>*zhoujinhua@ahmu.edu.cn</u>*
*<u>**shuhe.zhang@maastrichtuniversity.nl</u>;*

**Abstract**: The quality of quantitative differential phase contrast reconstruction (qDPC) can be severely degenerated by the mismatch of the background of two oblique illuminated images, yielding problematic phase recovery results. These background mismatches may result from illumination patterns, inhomogeneous media distribution, or other defocusing layers. In previous reports, the background is manually calibrated which is time-consuming, and unstable, since new calibrations are needed if any modification to the optical system was made. It is also impossible to calibrate the background from the defocusing layers, or for high dynamic observation as the background changes over time. To tackle the mismatch of background and increases the experimental robustness, we propose the Retinex-qDPC in which we use the images' edge features as data fidelity term yielding $L_2$-Retinex-qDPC and $L_1$-Retinex-qDPC for high background-robustness qDPC recontruction. The split Bregman method is used to solve the $L_1$-Retinex DPC. We compare both Retinex-qDPC models against state-of-the-art DPC reconstruction algorithms including total-variation regularized qDPC, and isotropic-qDPC using both simulated and experimental data. Results show that the Retinex qDPC can significantly improve the phase recovery quality by suppressing the impact of mismatch background. Within, the $L_1$-Retinex-qDPC is better than $L_2$-Retinex and other state-of-the-art DPC algorithms. In general, the Retinex-qDPC increases the experimental robustness against background illumination without any modification of the optical system, which will benefit all qDPC applications.

## 1. Introduction

From the first observation of phase contrast image in 1984 [1] to the development of quantitative differential phase contrast microscopy (qDPC) [2, 3], the DPC and the qDPC attracts many interests as it provides a non-interferometric phase retrieval approach and has been used for label-free and stain-free optical imaging of live biological specimens both in vitro and in vivo [4, 5].

    Under the weak object approximation [2], the forward model for qDPC data collection can be regarded as an image convolution process. The phase transfer function (PTF) converts the unmeasurable phase distribution of the sample into the differential phase contrast image i.e. the difference between two oblique illumination (OI) images under anti-symmetric illumination in opposite directions. Since the PTF is known from system parameters including illumination wavelength, parameters of the objective lens, and illumination patterns [2, 6-11], the inverse problem is described by a non-blind deconvolution where at least 4 OI images (2 differential

phase contrast images) or more are needed to increase the data redundancy for better phase reconstruction quality.

However, since no optical system is perfect, DPC raw images are always corrupted by noise signals, illumination fluctuations, or optical aberrations resulting in a degeneration of phase recovery results. In particular, the illumination fluctuations can severely degenerate qDPC reconstruction results since they introduce unexpected differential phase contrast components originating from the mismatch of backgrounds rather than the real phase of the sample.

Illumination fluctuations can be caused by unstable LED/LCD power supply, imperfect LED/LCD illumination quality, inhomogeneous media distribution, or interference shadows cast by suspension defocused layered samples and floating dust [12, 13], as well as vignetting effect [14]. It leads to severe problems in low-cost qDPC platforms, single-shot qDPC systems [15], and even in some traditional qDPC if the illumination is not carefully calibrated.

When the sample remains unchanged, a stable optical system for qDPC experiments should yield rather constant background components over time, which may be calibrated by taking images without a sample. However, in many studies, there is no automatic background correction or the background is manually measured and assumed to be unchanged even for high dynamic samples. Others suppress the background by modifying the illumination system [7, 9, 10], which increases the system's complicity and limits its applications.

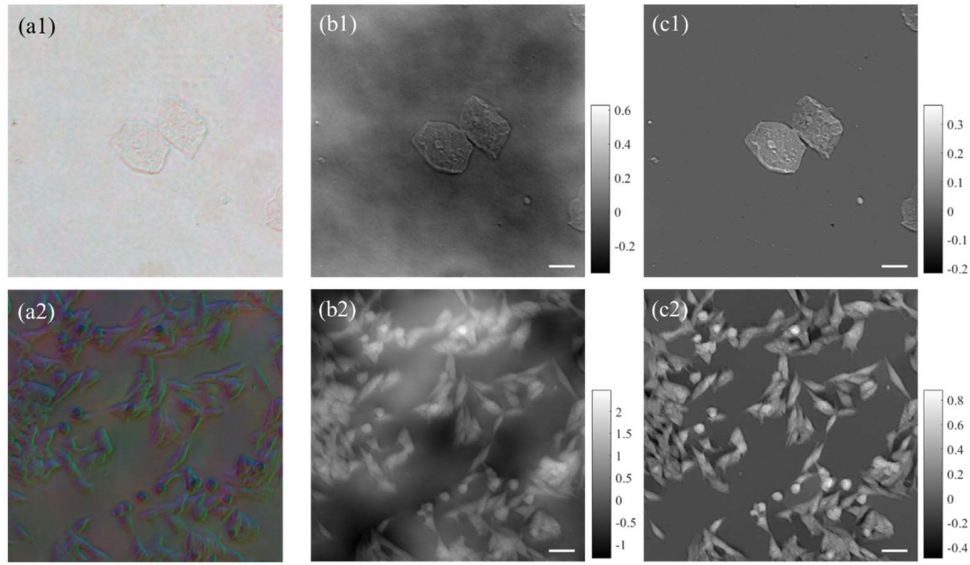

Fig. 1. Demonstration of Retinex DPC. (a1) and (a2) are input images for single-shot DPC for oral epithelium cells, and lung cancer cells. (b1) and (b2) are phase recovery results using a traditional DPC deconvolution algorithm with total variation regularization. (c1) and (c2) are reconstruction results using TV-$L_1$-Retinex DPC. Scale bar is 40 μm.

To suppress the impact of the background and further increase the quality of qDPC reconstruction, in this study we propose the Retinex-qDPC algorithm that uses the gradient of the qDPC images for phase reconstruction. This is pure mathematical approach that does not require any modification to the optical system of qDPC and can be applied to all qDPC experiments with different illumination layouts. Moreover, our Retinex-qDPC does not introduce additional modal parameters, which is superior to state-of-the-art qDPC reconstruction methods in both implementation efficiency, and phase reconstruction quality. Figure 1, (a1) to (c1), shows an example Retinex-qDPC reconstruction for a single-shot color multiplexed DPC of oral epithelium cells and (a2) to (c2) for lung cancer cells.

The rest of this manuscript is as follows: In section 2 we introduce the Retinex theory for DPC reconstruction algorithm and proposed to use the gradient of the image as the new data fidelity term for DPC deconvolution. We further propose two Retinex-qDPC models, the $L_2$-Retinex-qDPC and $L_1$-Retinex-qDPC to achieve background rectified phase reconstruction. In section 3, we formulate two new cost functions for the qDPC inverse problem for $L_2$-Retinex-qDPC and $L_1$-Retinex-qDPC, and two solutions for each case. Section 4 shows the experimental results for validation and application of the Retinex-qDPC and followed by a discussion in section 5 and a conclusion in section 6.

## 2. Retinex theory for DPC reconstruction

### 2.1 Maximum a posteriori estimation

DPC phase reconstruction is a typical Maximum a Posteriori (MAP) task [16]. Given a series of observed differential phase contrast (DPC) images $s_{dpc,n}$, $(n=1,2,\cdots N)$ corrupted according to the forward model of DPC and noise, the goal is to find the underlying phase distribution $\varphi$ that "best explains" the observations by maximizing the posterior probability:

$$\begin{aligned}\hat{\varphi} &= \arg\min_{\varphi} \prod_{n=1}^{N} \mathcal{P}(\varphi \mid s_{dpc,n}) \\ &= \arg\min_{\varphi} -\sum_{n=1}^{N} \log \mathcal{P}(s_{dpc,n} \mid \varphi) - \log \mathcal{P}(\varphi)\end{aligned}, \quad (1)$$

The first term in Eq. (1) denotes the probability of $s_{dpc,n}$ generated by the forward model of DPC with a given phase pattern $\varphi$, and the second term denotes the prior distribution $\mathcal{P}(\varphi)$ defining the probability distribution of the phase pattern.

The forward model of traditional DPC is a 2D convolution process [2, 6, 7, 17] that is given by:

$$s_{dpc,n} = h_{dpc,n} \otimes \varphi + \xi, \quad (2)$$

where $h_{dpc,n}$ is the $n$-th convolution kernel defined by $n$-th asymmetric illumination pupil and pupil function of the objective lens. $s_{dpc,n}$ is the corresponding $n$-th DPC image generated by the difference of images under asymmetric illumination patterns. Here, for example, $s_{dpc,n}$ is given by

$$s_{dpc,n} = \frac{I_{n,r} - I_{n,l}}{I_{n,r} + I_{n,l}}, \quad (3)$$

where $I_{n,r}$ and $I_{n,l}$ are captured OI images under anti-systematic oblique illumination from the right and left, respectively.

$\xi$ in Eq. (2) denotes the random noises imposed on the image. If it is assumed to be Gaussian distributed, Eq. (1) can be rewritten as

$$\hat{\boldsymbol{\varphi}} = \arg\min_{\varphi} \sum_{n=1}^{N} \left\| \boldsymbol{s}_{dpc,n} - \boldsymbol{H}_n \boldsymbol{\varphi} \right\|_2^2 + \alpha \cdot \Phi(\boldsymbol{\varphi}), \quad (4)$$

yielding the traditional DPC solver, where $\alpha\Phi(\boldsymbol{\varphi}) = -\log\mathcal{P}(\varphi)$ is the penalty/regularization function that limits the solution of Eq. (4), and $\Phi(\boldsymbol{\varphi}) = \|\boldsymbol{\varphi}\|_2^2$ is the widely-used $L_2$-DPC [2] (Tikhonov-DPC).

However, not all DPC images are obtained under ideal experimental conditions. Noise and uneven background illumination may degenerate the quality of the DPC reconstruction. Especially, uneven background illuminations can severely impact the reconstruction quality, as they introduce additional differential phase contrast structures to Eq. (3), that do not exist in an ideal condition.

To illustrate this, Fig. 2 shows a phantom DPC experiment. The raw images under oblique illumination from the left are shown in Fig. 2(b1) with uneven background, and in Fig. 2 (b2) without uneven background. Since the background is mismatched, the differential operation in Eq. (3) will introduce unexpected differential phase structures such as the white cloud in Fig. 2(e), which do not appear in the idea DPC conditions as shown in Fig. 2(d). Therefore, the final DPC deconvolution will enlarge the mismatched background signal, yielding problematic reconstruction results as shown in Fig. 2(f1).

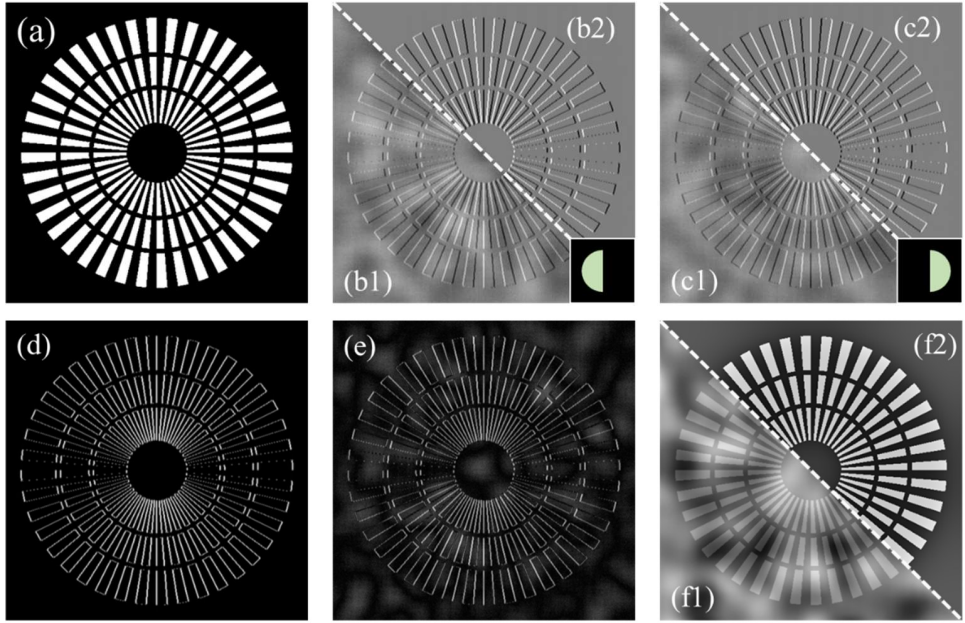

Fig. 2, phantom DPC experiment. (a) ground-truth phase pattern. (b1) and (b2) are a montage of captured images illuminated from the left. (b1) is further corrupted by uneven background. (c1) and (c2) are illuminated from the right. (d) and (e) are differential phase contrast images without and with the mismatch of background. (f1) and (f2) are reconstructed phase.

### 2.2 Threshold Retinex DPC

To tackle the uneven background illumination problem, we propose the Retinex-qDPC. We rewrite $I_{n,r}$ and $I_{n,l}$ as the pixel-wise superposition of structure components and background components which are

$$\begin{cases} I_{n,r} = S_{n,r} \circ (1-M) + B_{n,r} \circ M \\ I_{n,l} = S_{n,l} \circ (1-M) + B_{n,l} \circ M \end{cases}, \quad M = \begin{cases} 1, & \text{pixels} \in \text{background} \\ 0, & \text{else} \end{cases}, \quad (5)$$

where $S_{n,r}$ denotes the structural component and $B_{n,r}$ denotes the background component of $I_{n,r}$. $M$ is a binary sampling mask that divided pixels that belongs to the background component from the raw image $I_{n,r}$ and $I_{n,l}$. Moreover, $M$ is identical for both $I_{n,r}$, and $I_{n,l}$, since the background area is exactly matched pixels-by-pixels to each other so that they can be completely canceled out in Eq. (3) in the ideal condition. $\circ$ is the pixel-wise multiplication.

Since the multiplication in Eq. (5) and the division in Eq. (3) are performed in a pixel-wise manner, submitting Eq. (5) to Eq. (3) yielding

$$s_{dpc,n} = (1-M) \circ \frac{S_{n,r} - S_{n,l}}{S_{n,r} + S_{n,r}} + M \circ \frac{B_{n,r} - B_{n,l}}{B_{n,r} + B_{n,l}}. \tag{6}$$

Accordingly, the DPC image can be also divided into structural and background components, and most importantly, in an ideal condition, the second term in Eq. (6) is omitted as $B_{n,r} - B_{n,l} \equiv 0$, and the phase contrast structure purely originated from the first term (differential structure term):

$$s_{dpc,n}^{idea} = s_{dpc,n}\big|_{B_{n,r} \equiv B_{n,l}} = \frac{S_{n,r} - S_{n,l}}{S_{n,r} + S_{n,r}} = h_{dpc,n} \otimes \varphi. \tag{7}$$

When the background mismatch appears, the second term exists and impacts the reconstruction of $\varphi$.

Taking the image gradient to both sides of Eq. (6):

$$\nabla s_{dpc,n} = \nabla s_{dpc,n}^{idea} + \nabla B_n, \quad \nabla B_n = \nabla\left(\frac{B_{n,r} - B_{n,l}}{B_{n,r} + B_{n,l}}\right). \tag{8}$$

As we assume the background mismatch to be due to the uneven illumination, according to Retinex theory, $B_{n,r}$ and $B_{n,l}$ are spatially slow varying so that $\nabla B_n$ is relatively small [18-20], and $\nabla s_{dpc,n}^{idea}$ contains sharp details for DPC reconstruction. To preserve the $\nabla s_{dpc,n}^{idea}$, a threshold function is applied to $\nabla s_{dpc,n}$ [21], so that

$$\nabla s_{dpc,n}^{idea} \approx Threshold\left(\nabla s_{dpc,n}\right), \tag{9}$$

and a background rectified structure component $\nabla s_{dpc,n}^{idea}$ can be recovered by solving Eq. (9) for background-free DPC reconstruction results.

### 2.3 Variational Retinex DPC

The threshold operation in Eq. (9) increases the complicity of its implementation. Instead of solving the structural component in Eq. (9), we consider the variational Retinex framework [18, 22], which is mathematically equivalent to the threshold Retinex with curtain regularization [20], and directly embedded the Retinex in DPC reconstruction for solving of $\varphi$.

Submitting Eq. (8) in Eq. (2) and taking the gradient to both sides of the forward model Eq. (2) and ignoring the noise term we have

$$\nabla s_{dpc,n} = \nabla s_{dpc,n}^{idea} + \nabla B_n = h_{dpc,n} \otimes \nabla \varphi + \nabla B_n, \tag{10}$$

According to Eq. (10), we formulate our new data fidelity terms for DPC reconstruction which are

$$\hat{\boldsymbol{\varphi}} = \arg\min_{\varphi} \sum_{n=1}^{N} \left\| \nabla \boldsymbol{s}_{dpc,n} - \boldsymbol{H}_n \nabla \boldsymbol{\varphi} \right\|_2^2 + \alpha \cdot \Phi(\boldsymbol{\varphi}), \tag{11}$$

for $L_2$-Retinex DPC, and

$$\hat{\boldsymbol{\varphi}} = \arg\min_{\varphi} \sum_{n=1}^{N} \left\| \nabla \boldsymbol{s}_{dpc,n} - \boldsymbol{H}_n \nabla \boldsymbol{\varphi} \right\|_1 + \alpha \cdot \Phi(\boldsymbol{\varphi}), \tag{12}$$

for $L_1$-Retinex DPC. When the noise occurs, we are able to use the penalty function, such as the total variation, or its high order version to suppress the noise signals. The penalty function can be selected from one of the functions listed in Tab. 1 for different images prior, or using their combinations.

Tab. 1. List of DPC reconstruction solvers, denotes convoluting with Gaussian kernel.

| | Data fidelity term | + | Penalty term | Prior |
|---|---|---|---|---|
| Traditional DPC | $\sum_{n=1}^{N}\|s_{dpc,n} - H_n\varphi\|_2^2$ | | $\alpha\|\varphi\|_2^2$ | L2-norm/Weight decay |
| | | | $\alpha\|\nabla\varphi\|_1$ | Total variation (TV) [23] |
| L$_2$-Retinex DPC | $\sum_{n=1}^{N}\|\nabla s_{dpc,n} - H_n\nabla\varphi\|_2^2$ | + | $\alpha\|\nabla\nabla\varphi\|_1$ | High-order TV [24] |
| | | | $\alpha\|\nabla\varphi\|_2^2 + \beta\|\mathcal{G}\varphi\|_2^2$ | Isotropic-DPC [7] |
| L$_1$-Retinex DPC | $\sum_{n=1}^{N}\|\nabla s_{dpc,n} - H_n\nabla\varphi\|_1$ | | $\beta\sum_{n=1}^{N}\|H_n\varphi\|_0$ | Dark-field sparse prior |
| | | | Other penalties | |

The Retinex-DPC shares the same ideal of using image features for image restoration. In our case, the gradient operator in Eq. (11) and Eq. (12) extracts the edge features from the input image which is background-robustness as background contains no edges. In such a manner, the gradient operator can efficiently separate image structures from image background, and thus bypass the influence of background components.

From Eq. (10), minimizing $\|\nabla s_{dpc,n} - H_n\nabla\varphi\|_2^2$ is equivalent to minimizing $\|\nabla B_n\|_2^2$. Since $\nabla B_n$ is assumed to be of small values, using $\|\nabla s_{dpc,n} - H_n\nabla\varphi\|_2^2$ as the data fidelity function will diminish the impact of the background component as it gets smaller energy than $\|s_n - H_n\varphi\|_2^2$ since $\|B_n\|_2^2$ has a rather larger value than its gradient. While minimizing $\|\nabla s_{dpc,n} - H_n\nabla\varphi\|_1$ while getting sparser $\nabla B_n$ due to the feature of L$_1$-norm [19].

In the following section, we will take TV regularization as an example to deduce the solution to the backward problem of both L$_2$- and L$_1$-Retinex DPC in detail. Note that our Retinex DPC data fidelity can be applied to other penalty functions in a plug-and-play manner.

## 3. Backward problem

### 3.1 L$_2$-Retinex DPC

The L$_2$-Retinex DPC is quadratic and is easier to be solved than L$_1$-Retinex DPC. With TV regularization, the cost function for TV-L$_2$-Retinex DPC deconvolution is written as

$$E(\varphi) = \sum_{n=1}^{N}\|\nabla s_{dpc,n} - H_n\nabla\varphi\|_2^2 + \alpha \cdot \|\nabla\varphi\|_1, \quad (13)$$

Equation (13) can be solved by many methods including using half-quadratic splitting (HQS), iterative soft threshold, and split Bregman methods. Here we use the HQS method to solve the Eq. (13) [25].

Introducing two auxiliary variables $G = (G_x; G_y)$, $G_x = \nabla_x\varphi$, and $G_y = \nabla_y\varphi$, Eq. (13) is split into two sub-problems:

$$\begin{cases} E_1(\varphi) = \sum_{n=1}^{N}\|\nabla s_{dpc,n} - H_n\nabla\varphi\|_2^2 + \alpha_0\|G - \nabla\varphi\|_2^2 \\ E_2(G) = \alpha_0\|G - \nabla\varphi\|_2^2 + \alpha\|\nabla\varphi\|_1 \end{cases}, \quad (14)$$

where $\alpha_0$ is a sufficient large variable to ensure that $G = \nabla\varphi$. Both sub-problems have closed-form solutions. For $G$ sub-problem, the solution is

$$\begin{cases} \boldsymbol{G}_x = \dfrac{\nabla_x \boldsymbol{\varphi}}{\sqrt{(\nabla_x \boldsymbol{\varphi})^2 + (\nabla_y \boldsymbol{\varphi})^2}} \max\left[\sqrt{(\nabla_x \boldsymbol{\varphi})^2 + (\nabla_y \boldsymbol{\varphi})^2} - \alpha/\alpha_0, 0\right] \\ \boldsymbol{G}_y = \dfrac{\nabla_y \boldsymbol{\varphi}}{\sqrt{(\nabla_x \boldsymbol{\varphi})^2 + (\nabla_y \boldsymbol{\varphi})^2}} \max\left[\sqrt{(\nabla_x \boldsymbol{\varphi})^2 + (\nabla_y \boldsymbol{\varphi})^2} - \alpha/\alpha_0, 0\right] \end{cases}, \quad (15)$$

for isotropic TV-regularization, and

$$\begin{cases} \boldsymbol{G}_x = \text{sign}(\nabla_x \boldsymbol{\varphi}) \max(|\nabla_x \boldsymbol{\varphi}| - \alpha/\alpha_0, 0) \\ \boldsymbol{G}_y = \text{sign}(\nabla_y \boldsymbol{\varphi}) \max(|\nabla_y \boldsymbol{\varphi}| - \alpha/\alpha_0, 0) \end{cases}, \quad (16)$$

for anisotropic TV-regularization. Since $E_1$ is pure quadratic, the solution of $\boldsymbol{\varphi}$ can be found by setting $\partial E_1 / \partial \boldsymbol{\varphi} = 0$, yielding

$$\boldsymbol{\varphi} = \frac{\left(\nabla_x^T \nabla_x + \nabla_y^T \nabla_y\right) \sum_{n=1}^{N} \boldsymbol{H}_n^T \boldsymbol{s}_{dpc,n} + \alpha_0 \left(\nabla_x^T \boldsymbol{G}_x + \nabla_y^T \boldsymbol{G}_y\right)}{\left(\nabla_x^T \nabla_x + \nabla_y^T \nabla_y\right)\left(\sum_{n=1}^{N} \boldsymbol{H}_n^T \boldsymbol{H}_n + \alpha_0\right) + \eta}. \quad (17)$$

Eq. (17) can be calculated using Fourier transform since both $\nabla$ operator and $\boldsymbol{H}_n$ denote convolution process. $\eta$ is a very small value to avoid dividing by zeros, which is mathematically equivalent to L$_2$-norm regularization, and $\eta = 10^{-6}$ throughout the study. The algorithm for solving TV L$_2$-Retinex DPC is summarized in the following list.

---

**Algorithm 1**: TV L$_2$-Retinex DPC

**Input**: $\boldsymbol{s}_{dpc,n}$, $\boldsymbol{H}_n$, $\alpha$, $\alpha_0 = \alpha$, $\alpha_{\max} = 10^5$, initializing $\boldsymbol{\varphi} = 0$

**While** $\alpha_0 < \alpha_{\max}$ **do**

    Calculate $\nabla_x \boldsymbol{\varphi}$ and $\nabla_y \boldsymbol{\varphi}$;
    Solving $\boldsymbol{G}$ - sub problem using Eq. (15)/(16) for iso/aniso-tropic TV regularization;
    Solving $\boldsymbol{\varphi}$ - sub problem using Eq. (17);
    $\alpha_0 = 2\alpha_0$;

**End**
Output: Background rectified $\boldsymbol{\varphi}$

---

The first term in Eq. (14) can be also regarded as a Least Squares procedure, and the L$_2$-norm is known as ridge regression, which terms to minimize the value of $\|\nabla B_n\|_2^2$, and smooths the background $B_n$. However, in an idea DPC condition or microscopy condition, the $B_n$ should not only be smoothed but also be constant since a uniformly distributed background is preferred. The L$_2$ Retinex will keep some small value of $\|\nabla B_n\|_2^2$ resulting in gentle variance of background $B_n$, which is sometimes not as desired.

### 3.2 L$_1$-Retinex DPC

If the background is more evenly distributed, its gradient should be sparser. We introduce L$_1$-Retinex DPC by minimizing the L$_1$-norm of the data fidelity term. With TV regularization, the cost function for TV- L$_1$-Retinex DPC is given as

$$E = \gamma \cdot \sum_{n=1}^{N} \left\| \nabla s_{dpc,n} - H_n \nabla \varphi \right\|_1 + \alpha \cdot \left\| \nabla \varphi \right\|_1, \tag{18}$$

The L$_1$-norm is able to make the $\nabla s_{dpc,n} - H_n \nabla \varphi$ sparser than L$_2$-norm. Thus, with a high probability we will have a DPC reconstruction results with a constant background.

The data fidelity term in Eq. (18) can be expanded into $2N$ L$_1$-norm terms. Similar to TV regularization, the $2N$ L$_1$-norm can be solved in isotropic and anisotropic behaviors, here we will treat only the isotropic case in detail since the treatment for the anisotropic case is completely analogous. Since there are two L$_1$-norm terms in Eq. (18), we use the split Bregman method for fast and better implementation [22].

Introducing $2N + 2$ variables $\psi_{n,x} = \nabla_x s_{dpc,n} - H_n \nabla_x \varphi$, $\psi_{n,y} = \nabla_y s_{dpc,n} - H_n \nabla_y \varphi$, and $G_x = \nabla_x \varphi$, and $G_y = \nabla_y \varphi$, the split Bregman iteration for Eq. (18) at k-th iteration is given by

$$\varphi^{k+1} = \arg\min \ \gamma_0 \sum_{\upsilon=x,y} \sum_{n=1}^{N} \left\| b_{n,\upsilon}^k - \psi_{n,\upsilon}^k + H_n \nabla_\upsilon \varphi - \nabla_\upsilon s_{dpc,n} \right\|_2^2 \\ + \alpha_0 \sum_{\upsilon=x,y} \left\| G_\upsilon - g_\upsilon - \nabla_\upsilon \varphi \right\|_2^2, \tag{19-1}$$

$$\psi_{n,\upsilon}^{k+1} = \arg\min \ \gamma_0 \sum_{\upsilon=x,y} \sum_{n=1}^{N} \left\| b_{n,\upsilon}^k - \psi_{n,\upsilon}^k + H_n \nabla_\upsilon \varphi - \nabla_\upsilon s_{dpc,n} \right\|_2^2 + \gamma \left\| \psi_{n,\upsilon} \right\|_1, \tag{19-2}$$

$$G_\upsilon^{k+1} = \arg\min \ \alpha_0 \sum_{\upsilon=x,y} \left\| G_\upsilon - \nabla_\upsilon \varphi^{k+1} - g_{\upsilon,n} \right\|_2^2 + \alpha \left\| G_\upsilon \right\|_1, \tag{19-3}$$

$$b_{n,\upsilon}^{k+1} = b_{n,\upsilon}^k + \left( H_n \nabla_\upsilon \varphi^{k+1} - \nabla_\upsilon s_{dpc,n} \right) - \psi_{n,\upsilon}^{k+1}, \tag{19-4}$$

$$g_\upsilon^{k+1} = g_\upsilon^k + \nabla_\upsilon \varphi^{k+1} - G_\upsilon^{k+1}, \tag{19-5}$$

where $\gamma_0$ and $\alpha_0$ are additional parameters introduced by the split Bregman method, and $\gamma_0 = 1$ and $\alpha_0 = 1$ by default. $b_{n,\upsilon}^k$ and $g_\upsilon^k$ are intermedia variable in split Bregman method.

The Eq. (19-1) is pure quadratic, and the closed-form solution is

$$\varphi^{k+1} = \frac{\gamma_0 \sum_{\upsilon=\{x,y\}} \sum_{n=1}^{N} \left[ H_n^T \nabla_\upsilon^T \nabla_\upsilon s_{dpc,n} + H_n^T \nabla_\upsilon^T \left( \psi_{n,\upsilon}^k - b_{n,\upsilon}^k \right) \right] + \alpha_0 \sum_{\upsilon=\{x,y\}} \nabla_\upsilon^T \left( G_\upsilon^k - g_\upsilon^k \right)}{\left( \nabla_x^T \nabla_x + \nabla_y^T \nabla_y \right) \left( \gamma_0 \sum_{n=1}^{N} H_n^T H_n + \alpha_0 \right) + \eta}. \tag{20}$$

The closed-form solution of Eq. (19-2) for isotropic Retinex is

$$\begin{cases} \psi_{n,\upsilon}^{k+1} = \dfrac{H_n \nabla_\upsilon \varphi^{k+1} - \nabla_\upsilon s_{dpc,n} + b_{n,\upsilon}^k}{\Omega} \max\left( \Omega - \gamma / \gamma_0 \right) \\ \Omega = \sqrt{\sum_{\upsilon=\{x,y\}} \sum_{n=1}^{N} \left( H_n \nabla_\upsilon \varphi^{k+1} - \nabla_\upsilon s_{dpc,n} + b_{n,\upsilon}^k \right)^2} \end{cases}. \tag{21}$$

The closed-form solution of $G$ is given by Eq. (15) and Eq. (16) by replacing $\nabla_\upsilon \varphi$ with $\nabla_\upsilon \varphi + g_\upsilon$.

The split Bregman method is performed iteratively, and empirically, 40 iterations will lead to stable convergent. The algorithm for solving TV L$_1$-Retinex DPC is summarized in the following list. The soft-threshold in Eq. (21) can be also applied to individual $\psi_{n,\upsilon}^{k+1}$ yielding anisotropic $L_1$-Retinex. In the following experiment, we show only the results of isotropic $L_1$-Retinex.

---

**Algorithm 1**: TV $L_1$-Retinex DPC
---

**Input**: $s_{dpc,n}$, $\boldsymbol{H}_n$, $\alpha$, $\alpha_0 = 1$, $\gamma_0 = 1$, $\gamma$, $k = 0$
  initializing $\boldsymbol{\varphi}^0 = \boldsymbol{b}_{n,\upsilon}^0 = \boldsymbol{g}_\upsilon^0 = \boldsymbol{\psi}_{n,\upsilon}^0 = \boldsymbol{G}_\upsilon^0 = 0$

**While** $k < \max\_\text{iterations}$ **do**

  Solving $\boldsymbol{\varphi}^{k+1}$ using Eq. (20);

  Solving $\boldsymbol{\psi}_{n,\upsilon}^{k+1}$ using Eq. (21);

  Solving $\boldsymbol{G}_\upsilon^{k+1}$ using Eq. (15) or Eq. (16);

  Update $\boldsymbol{b}_{n,\upsilon}^{k+1}$ and $\boldsymbol{g}_\upsilon^{k+1}$ using Eq. (19-4) and Eq. (19-5);

  $k = k + 1$;

**End**

Output: Background rectified $\boldsymbol{\varphi}$

---

## 4. Experimental results

### 4.1 Phantom phase sample

First, we perform DPC reconstruction using a phantom phase sample as shown in Fig. 2(a) where the phase is ranged in [0, 2] rad. System parameters are $\lambda = 0.530$ μm, $NA = 0.3$, magnification is ×10 and the pixel size of the camera is 3.46 μm. We add random uneven background to the simulated raw image. The random Gaussian noise is also added and the signal-to-noise ratio is 20 db.

For model implementation, the penalty parameter $\alpha$ for TV-regularization is adaptively determined using the adaptive noise sensor:

$$\alpha = \frac{1}{20}\sqrt{\frac{\pi}{2}}\frac{1}{N}\frac{1}{WH}\sum_{n=1}^{N}\sum_{x=1}^{W}\sum_{y=1}^{H}\left|s_n(x,y)\otimes\mathcal{L}\right|, \qquad (22)$$

where $\mathcal{L} = [-1, 2, -1; 2, -4, 2; -1, 2, -1]$ is the Laplacian operator. $\beta = 2\alpha$ for Iso-DPC. $\gamma = 0.1$ for TV-$L_1$-Retinex DPC.

Fig. 3(a) shows reconstruction results for TV-DPC, where the image is severely degenerated by the background signal. The phase pattern is merged in the 'white clouds' due to the mismatch of background. Fig. 3 (b) shows Retinex-DPC results for TV-$L_2$-Retinex and Fig. 3 (c) for TV-$L_1$-Retinex. The background signal is significantly suppressed, and the recovered phase pattern can be clearly observed. Moreover, the background in TV-$L_1$-Retinex is more uniform than that in TV-$L_2$-Retinex, due to the sparse promotion of $L_1$-norm.

Quantitative phase measures along the yellow curves are plotted in Fig. 3 (d), where the curves (green and blue) for Retinex DPC are close to the ground-truth value, denoting high-data fidelity phase reconstruction when the uneven background problem appears. In contrast, results for normal TV-DPC (red curve) vary from +3 rad to -2 rad, due to the impact of background signals. Both two Retinex methods output high data fidelity reconstruction results as the curves approach the ground-truth one (black), with only constant phase shifts.

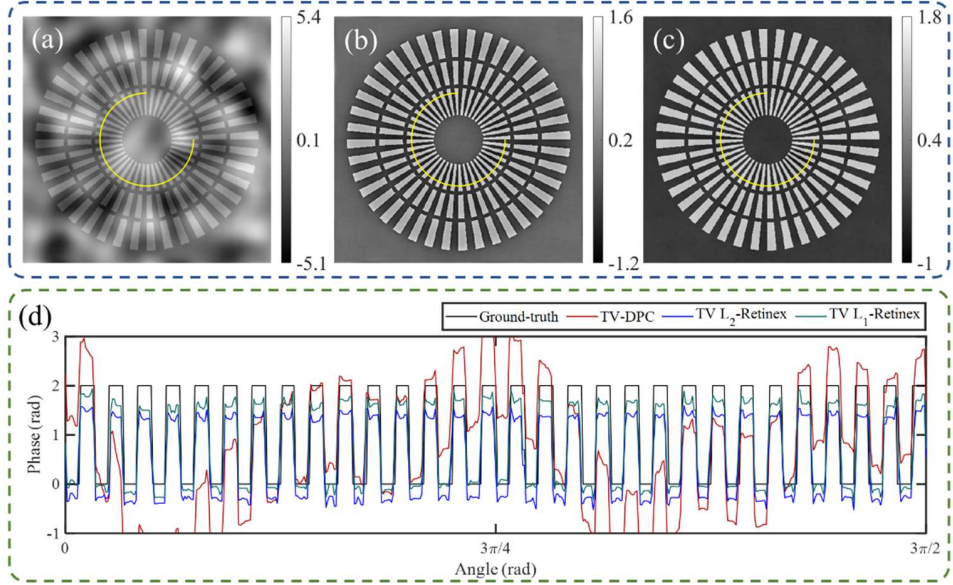

Fig. 3. Phantom DPC experiments. The input DPC images are shown in Fig. 2. (a) Results for TV-DPC. (b) Results for TV-L$_2$-Retinex DPC. (c) Results for TV-L$_1$-Retinex DPC. (d) quantitative phase profiles along the yellow arcs.

To get rid of the constant phase shifts and quantitatively measure the quality of the reconstructed phase in the simulation study, we use the regressed phase SNR (rpSNR) between the ground truth $\tilde{\varphi}$ and calculated $\varphi$ by solving the following 1D optimization problem:

$$rpSNR(\varphi, \tilde{\varphi}) = \max_{c \in R} \; 10\log\left(\frac{\|\tilde{\varphi}\|_2^2}{\|\tilde{\varphi} - (\varphi + c)\|_2^2}\right), \tag{23}$$

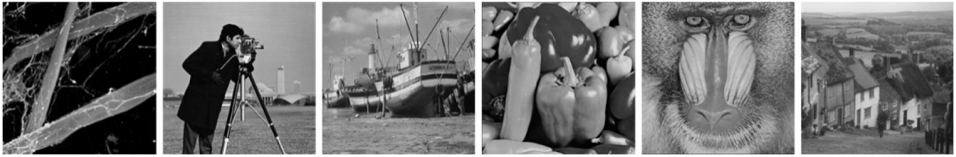

Fig. 4. Ground-truth spatial phase maps used in the simulations. From left to right, they are referred to as (1) CIL38921, (2) Cameraman, (3) Boat, (4) Peppers, (5) Baboon, and (6) Goldhill.

We further performed a massive simulation study on different input ground-truth phase patterns (see Fig. 4). All the ground-truth phase patterns are of size (600 by 600) pixels and have values in the range of [0, 1] rads. The experimental parameters are identical to that for Fig. 3. For each phase pattern, we generate the simulated DPC image using forward model:

$$\begin{cases} \boldsymbol{s}_{\text{idea},n} = \boldsymbol{H}_n \tilde{\boldsymbol{\varphi}} \\ \boldsymbol{s}_{\text{degraded},n} = \boldsymbol{s}_{\text{idea},n} + \boldsymbol{\xi} + A\left[\max(\boldsymbol{s}_{\text{idea},n}) - \min(\boldsymbol{s}_{\text{idea},n})\right]\boldsymbol{B} \end{cases}, \tag{24}$$

where $\boldsymbol{\xi}$ is the Gaussian noises generated according to the given SNR. $\boldsymbol{B}$ is the amplitude normalized background mismatch, and $A$ controls the amplitude of background mismatch.

The strength of Gaussian noises can be adjusted to meet certain SNR levels. For each phase pattern, we add two levels of Gaussian noises including 20db denoting large noises condition, and 40db denoting small noises condition. In each level of Gaussian noise, we further add 4 groups of background signals where $A=\{0,0.25,0.5,0.75\}$ to test the algorithm performance on different strengths of the background. We repeat the simulation experiment 10 times for each group for reliable comparison. The performance of our Retinex DPC is compared against to state-of-the-art DPC reconstruction algorithms including TV-DPC, Iso-DPC. The parameters for TV-regularization are adaptively determined using Eq. (22). The parameter $\gamma=0.1$ when $A=0$ and $0.25$, and $\gamma=0.01$ for $A=0.5$, and $\gamma=0.002$ for $A=0.75$.

**Table 2. Phase reconstruction performance of the algorithm compared in the simulations**

| Input SNR (db) | | Strength $A$ | Methods ( rpSNR db) | | | |
|---|---|---|---|---|---|---|
| | | | TV-DPC | Iso-DPC | TV-L$_2$-Retinex | TV-L$_1$-Retinex |
| CIL38921 | 20 | 0.00 | **31.27 ± 0.32** | 13.53 ± 0.03 | 10.43 ± 0.21 | 11.59 ± 0.16 |
| | | 0.25 | -3.23 ± 2.65 | 8.83 ± 0.61 | 10.50 ± 0.21 | **11.04 ± 0.15** |
| | | 0.50 | -18.75 ± 4.43 | 1.23 ± 1.03 | 10.31 ± 0.34 | **10.97 ± 0.10** |
| | | 0.75 | -26.29 ± 2.00 | -5.42 ± 0.89 | 10.09 ± 0.22 | **10.82 ± 0.15** |
| | 40 | 0.00 | **39.09 ± 0.04** | 23.43 ± 0.01 | 15.81 ± 0.03 | 21.29 ± 0.10 |
| | | 0.25 | -5.23 ± 2.50 | 3.26 ± 1.43 | **15.4 ± 0.29** | 14.98 ± 0.29 |
| | | 0.50 | -18.92 ± 1.80 | -9.65 ± 1.58 | **13.6 ± 0.36** | 9.39 ± 0.05 |
| | | 0.75 | -28.18 ± 3.21 | -17.50 ± 1.59 | **11.57 ± 0.89** | 9.16 ± 0.08 |
| Cameraman | 20 | 0.00 | **42.90 ± 0.47** | 21.01 ± 0.01 | 22.75 ± 0.21 | 24.60 ± 0.26 |
| | | 0.25 | 7.80 ± 2.80 | 17.14 ± 0.55 | 22.72 ± 0.29 | **24.58 ± 0.30** |
| | | 0.50 | -7.49 ± 3.78 | 10.47 ± 0.97 | 22.58 ± 0.35 | **24.48 ± 0.23** |
| | | 0.75 | -16.17 ± 2.27 | 4.51 ± 0.74 | 21.94 ± 0.54 | **24.13 ± 0.22** |
| | 40 | 0.00 | **48.13 ± 0.04** | 26.60 ± 0.01 | 28.16 ± 0.04 | 26.80 ± 0.06 |
| | | 0.25 | 4.42 ± 2.28 | 11.76 ± 1.25 | **27.09 ± 0.23** | 26.00 ± 0.21 |
| | | 0.50 | -9.80 ± 3.30 | 0.37 ± 1.11 | **25.22 ± 0.38** | 20.20 ± 0.08 |
| | | 0.75 | -18.97 ± 3.40 | -8.48 ± 1.67 | **22.50 ± 0.57** | 19.89 ± 0.07 |
| Boat | 20 | 0.00 | **44.58 ± 0.35** | 30.19 ± 0.03 | 26.40 ± 0.32 | 27.14 ± 0.18 |
| | | 0.25 | 6.88 ± 1.93 | 23.00 ± 1.00 | 26.24 ± 0.24 | **26.57 ± 0.27** |
| | | 0.50 | -7.09 ± 3.57 | 12.64 ± 1.24 | 25.96 ± 0.23 | **26.58 ± 0.21** |
| | | 0.75 | -15.32 ± 2.88 | 6.37 ± 0.93 | 25.82 ± 0.33 | **26.44 ± 0.33** |
| | 40 | 0.00 | **53.11 ± 0.06** | 36.30 ± 0.01 | 29.87 ± 0.03 | 33.56 ± 0.10 |
| | | 0.25 | 2.91 ± 2.57 | 12.97 ± 1.48 | **29.42 ± 0.30** | 29.24 ± 0.31 |
| | | 0.50 | -8.24 ± 2.90 | 1.88 ± 1.97 | **27.99 ± 0.48** | 25.64 ± 0.06 |
| | | 0.75 | -17.39 ± 2.98 | -7.14 ± 1.33 | **25.57 ± 0.57** | 25.48 ± 0.08 |
| Peppers | 20 | 0.00 | **37.57 ± 0.25** | 24.51 ± 0.04 | 20.69 ± 0.16 | 22.37 ± 0.14 |
| | | 0.25 | 4.71 ± 2.30 | 19.00 ± 0.72 | 20.64 ± 0.16 | **21.84 ± 0.14** |
| | | 0.50 | -10.26 ± 3.21 | 10.88 ± 0.72 | 20.26 ± 0.30 | **21.65 ± 0.15** |
| | | 0.75 | -18.03 ± 2.78 | 4.00 ± 1.31 | 19.85 ± 0.39 | **21.50 ± 0.15** |
| | 40 | 0.00 | **47.12 ± 0.04** | 33.93 ± 0.01 | 25.97 ± 0.02 | 31.40 ± 0.10 |
| | | 0.25 | 1.38 ± 2.88 | 12.14 ± 1.65 | **25.58 ± 0.17** | 25.19 ± 0.16 |
| | | 0.50 | -10.35 ± 2.84 | 0.32 ± 1.68 | **24.47 ± 0.37** | 19.10 ± 0.06 |
| | | 0.75 | -19.56 ± 2.74 | -8.60 ± 1.39 | **22.45 ± 0.42** | 18.80 ± 0.07 |
| Baboon | 20 | 0.00 | **44.25 ± 0.27** | 34.38 ± 0.05 | 29.71 ± 0.29 | 30.49 ± 0.21 |
| | | 0.25 | 8.11 ± 1.96 | 26.17 ± 0.87 | 29.82 ± 0.13 | **30.07 ± 0.15** |
| | | 0.50 | -3.54 ± 2.85 | 16.59 ± 1.00 | 29.26 ± 0.23 | **29.90 ± 0.19** |
| | | 0.75 | -14.23 ± 2.04 | 8.42 ± 1.11 | 28.77 ± 0.49 | **29.64 ± 0.36** |
| | 40 | 0.00 | **51.21 ± 0.03** | 42.25 ± 0.01 | 32.53 ± 0.04 | 38.98 ± 0.13 |
| | | 0.25 | 7.86 ± 2.23 | 18.50 ± 1.56 | 31.80 ± 0.29 | **32.15 ± 0.26** |
| | | 0.50 | -7.55 ± 3.52 | 5.63 ± 1.79 | **29.66 ± 0.91** | 28.07 ± 0.05 |
| | | 0.75 | -13.85 ± 4.24 | -1.93 ± 1.45 | 26.66 ± 0.79 | **27.99 ± 0.04** |
| Goldhill | 20 | 0.00 | **37.08 ± 0.35** | 21.58 ± 0.02 | 19.41 ± 0.19 | 19.99 ± 0.22 |
| | | 0.25 | 3.11 ± 3.71 | 16.42 ± 0.79 | 19.20 ± 0.14 | **19.45 ± 0.19** |
| | | 0.50 | -10.42 ± 3.10 | 9.08 ± 0.99 | 19.09 ± 0.29 | **19.46 ± 0.19** |
| | | 0.75 | -17.44 ± 2.60 | 2.86 ± 1.26 | 18.79 ± 0.48 | **19.38 ± 0.21** |
| | 40 | 0.00 | **46.16 ± 0.05** | 28.44 ± 0.01 | 21.92 ± 0.03 | 24.96 ± 0.08 |
| | | 0.25 | 1.92 ± 2.39 | 11.02 ± 1.57 | **21.61 ± 0.18** | 19.04 ± 0.74 |
| | | 0.50 | -12.90 ± 2.46 | -1.60 ± 1.27 | **20.62 ± 0.23** | 20.24 ± 0.18 |
| | | 0.75 | -21.49 ± 2.43 | -11.05 ± 1.51 | 18.94 ± 0.48 | **19.47 ± 0.29** |

From the simulated experiments, we found that when no background mismatch occurs, the traditional TV-DPC gain high rpSNR scores, denoting high-quality phase reconstruction results. However, when a background mismatch occurs, even a slight mismatch of background will significantly degrade the rpSNR for TV-DPC, for example, the rpSNR score for the Boat image is 44.58 when $A = 0$, while decreases to 6.88 for $A = 0.25$. A similar problem happens to the Iso-DPC algorithm as some of the rpSNR is even smaller than 0. Figure 5 shows one of the examples of phase reconstruction results when SNR = 40, $A = 0.25$, where both TV-DPC and Iso-DPC are severely corrupted by the 'white-fog' effect.

On the contrary, Retinex DPC are more robust to the background mismatch, as they generate stable rpSNR scores for different values of $A$. Especially, for large $A$, the results for TV-DPC and Iso-DPC already collapse as the rpSNR scores are less than zero, but both Retinex DPC generates rather high and stable rpSNR scores.

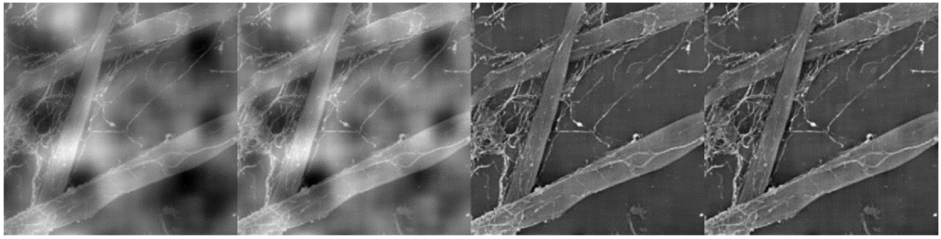

Fig. 5, Comparison of phase reconstruction results. Images from left to right are results for TV-DPC, Iso-DPC, TV-$L_2$-Retinex DPC, and TV-$L_1$-Retinex DPC.

### 4.2 Quantitative phase target

We validated our Retinex DPC among quantitative phase targets (QPT) (see Fig. 6). We used a focal star, 100 nm in height, and the refractive index was 1.52. The QPT was imaged by an objective lens with $NA = 0.3$, $\times 10$. The pixel size of the camera is 6.5 μm. We illuminate the QPT using a ring-shaped LED array and the illumination pupil is also a ring shape whose outer radius was $NA/\lambda$, and inner radius was $0.9\ NA/\lambda$.

We collected 4 OI images under oblique illumination corresponding to top, bottom, right, and left illumination, obtaining 2 groups of DPC images that are used for phase reconstruction. We compared our reconstruction results with those using TV-DPC and Iso-DPC.

For visual assessments, as shown in Fig. 6 (c), when the phase is reconstructed by TV-DPC, the phase pattern has a 'white-fog' effect due to the mismatch of background. The results for Iso-DPC are shown in Fig. 6(d), which is better than that of TV-DPC as the background in the recovered phase pattern is more uniformly distributed than that in Fig. 6(c), and the 'white-fog' effect is suppressed. The TV-$L_2$-Retinex DPC is shown in Fig. 6 (e), where the background signal is significantly suppressed. No 'white-fog' effect appears. With $L_1$-norm regularization, the TV-$L_1$-Retinex DPC ($\gamma = 1$) obtains more even background, and higher image contrast as the grayscale of the phase structure is larger than that in Fig. 6 (d) and 6 (e).

For quantitative assessments, the phase profiles are plotted along the yellow and orange arcs in Fig. 6(b), where the Iso-DPC, and both of the $L_2$-Retinex DPC maintain the data fidelity of the phase reconstruction, as their phase values approach the ground-truth with only constants background phase-shift. While the phase-shift for TV-DPC is large due to the impact of background mismatch. Since the DPC measures the phase difference between the sample and surrounding media, the mismatch of background will cause unexpected background phase differences and further degenerate the phase recovery results.

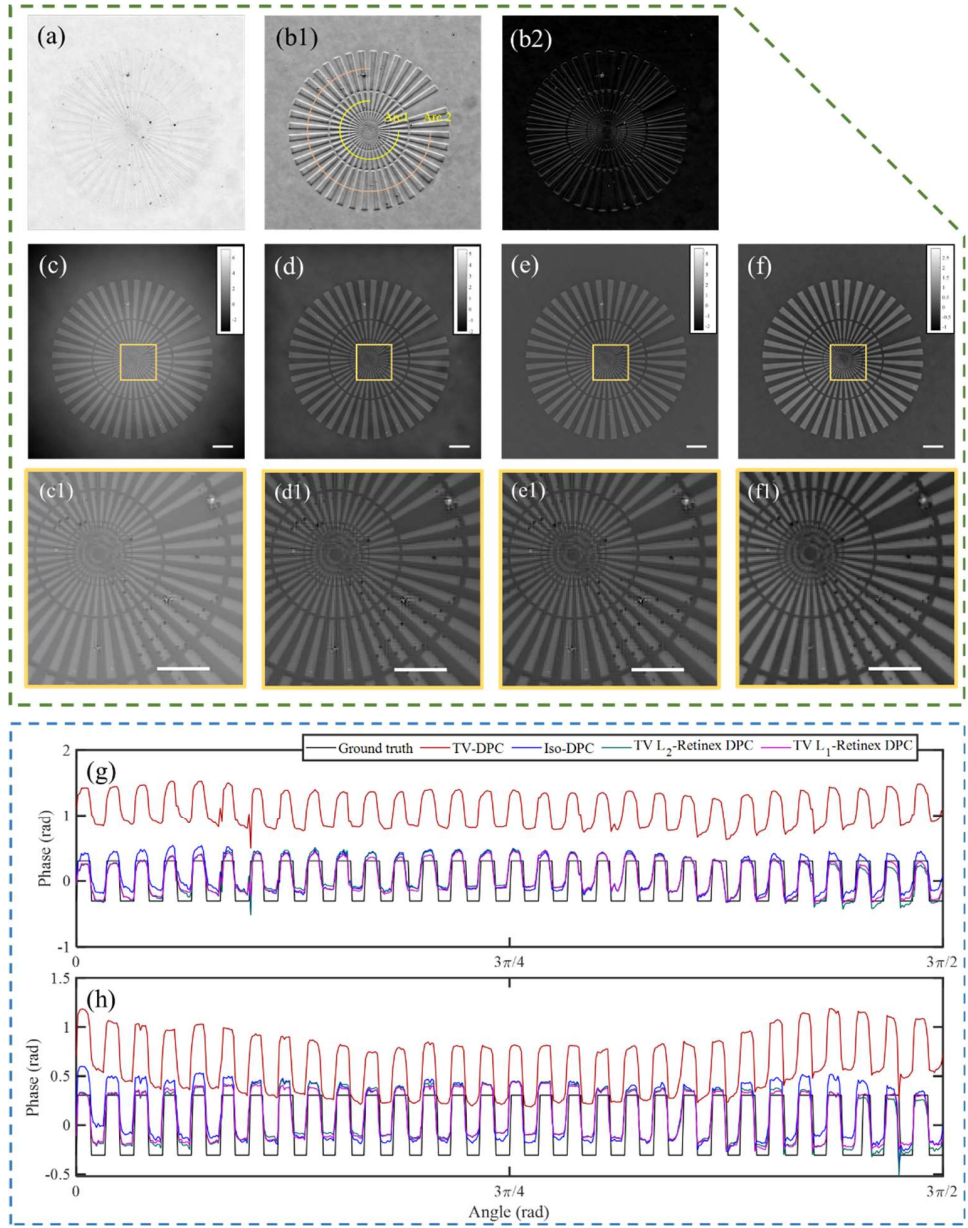

Fig. 6. Reconstruction results for the focal-star pattern (a) Bright-field image. (b1) input image under oblique illumination. (b2) DPC image. (c) TV-DPC. (d) Iso-DPCP. (e) and (f) are results of TV-$L_2$-Retinex DPC, and TV-$L_1$-Retinex DPC. The scale bar is 50 for raw images, and the scale bar is 20 μm for the zoomed-in images (c1 – f1). (g) and (h) are plots of phase profiles along the yellow arc and the orange arc in (b).

We performed another group of experiments for the USAF phase target. The QPT is imaged by an objective lens with $NA = 0.75$, ×40. The pixel size of the camera was 5.86 μm (SUA231GM-T, MindVision, China). $\gamma = 0.1$, All other parameters remained unchanged.

As shown in Fig. 7(a), the problematic background is even prominent in large magnification. A large exposure time is needed so that more noises appear. Due to the mismatch of background, the phase recovery results for both TV-DPC and Iso-DPC are severely degraded as shown in Fig. 7(c) and 7(d).

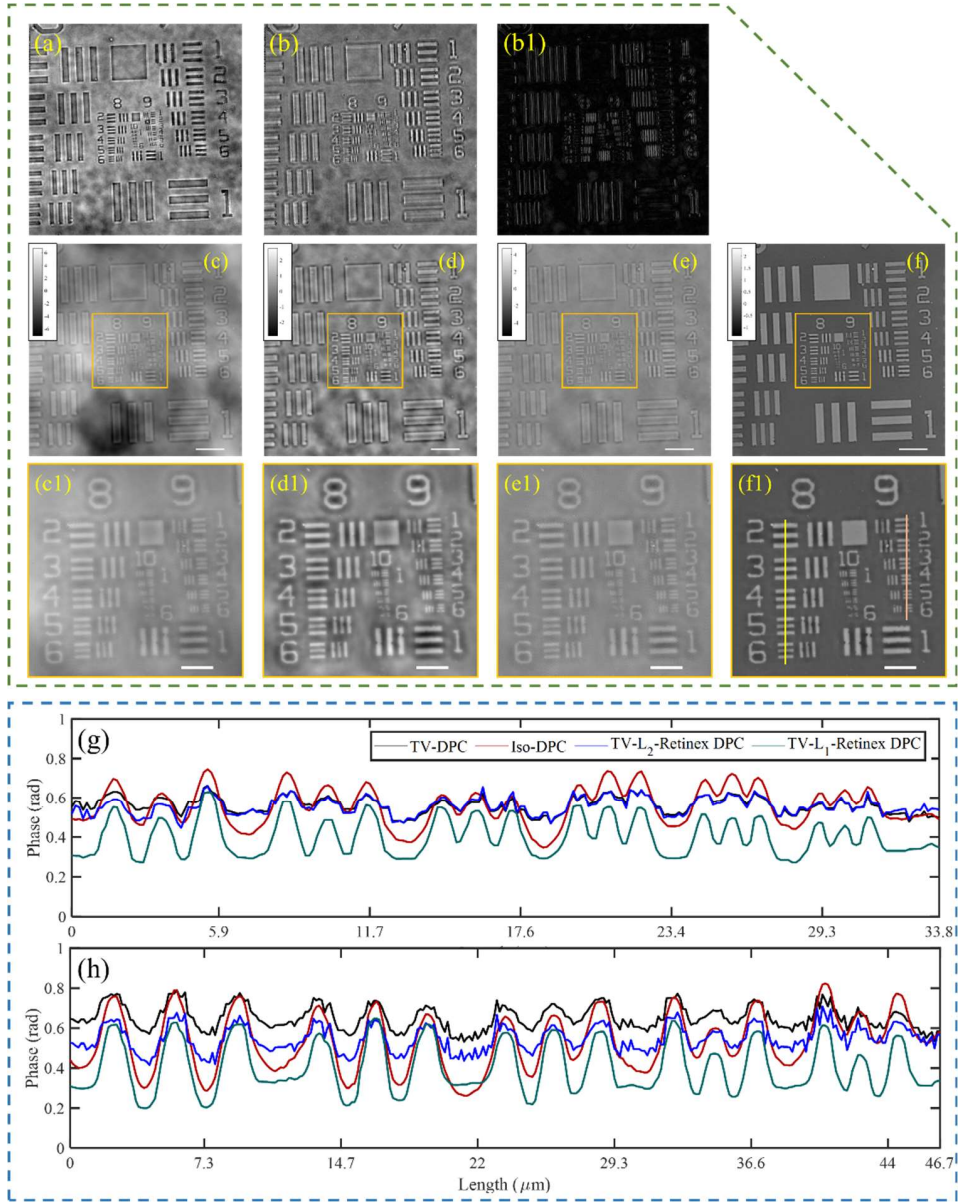

Fig. 7. Reconstruction results for USAF pattern (a) Bright-field image. (b) Input image under oblique illumination. (c) TV-DPC. (d) Iso-DPCP. (e) and (f) are results of TV-$L_2$-Retinex DPC, and TV-$L_1$-Retinex DPC. The scale bar is 50 for raw image, and is 20 μm for the zoomed-in images. (g) and (h) are plot of phase profile along the yellow and the orange line in (f).

Fig. 7(e) shows recovery result for TV-$L_2$-Retinex DPC. Although the background noise is suppressed, there are still uneven background components as can be seen in the zoomed-in image. In this case, the TV-$L_1$-Retinex DPC shows its superiority in the background rectifying. As shown in Fig. 7(f), the uneven background is completely removed, and the entire background in 7(f) is evenly distributed. The visual quality of Fig. 7(f) is better than the other three images, as no 'white-fog' effect exists.

The quantitative phase profile along the yellow and orange lines in Fig. 7(f) are plotted in Fig. 7(g) and 7(h). Accordingly, the TV-$L_1$-Retinex DPC shows its ability to background rectification, and noise suppression, which improves the quality of DPC phase reconstruction.

### 4.3 Full-field background rectifying

In certain DPC applications where the sample is of high dynamic, the exposure time may need to be decreased in order to get high frame rates. As such, the captured raw image may suffer from a low signal-to-noise ratio and be corrupted by problematic background.

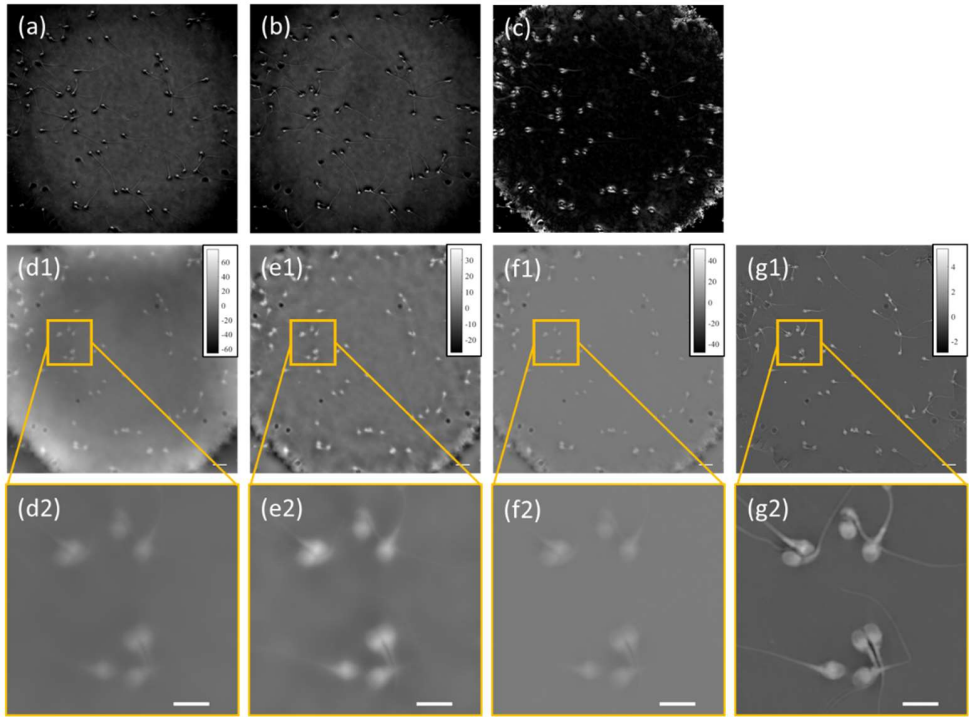

Fig. 8. DPC experiment of sperms sample with low intensity. (a) and (b) are raw image under oblique illuminations. (c) is the absolute value of DPC image of (a) and (b). (d1) to (g1) are reconstruction results using TV-DPC, Iso-DPC, TV-$L_2$-Retinex DPC, and TV-$L_1$-Retinex DPC, respectively. The scale bar is 20 μm (d2) to (g2) are enlarged part in the yellow boxes. The scale bar is 10 μm.

For example, Fig. 8(a) and 8(b) show two phase-contrast images for sperm samples illuminated from top and bottom, where $NA = 0.75$, ×40. The pixel size of the camera was 6.4 μm. The insufficient intensity, as well as vignetting effect, degenerates the image quality, especially at the edge of the field of view. The DPC image in Fig 8(c) presents the mismatch background problem.

In this case, the traditional DPC algorithms fail to recover the correct phase distribution as shown in Fig. 8(d1) for TV-DPC, and Fig. 8(e1) for Iso-DPC, where the image is corrupted by the heavy 'white-fog' effect due to the problematic backgrounds. The heads and tails of the cells can hardly be observed. Our TV-$L_2$-Retinex can suppress the background as shown in Fig. 8(f1) and 8(f2), but the white-fog' effect is still prominent. While our TV-$L_1$-Retinex DPC ($\gamma = 0.001$) can correct background among the entire field of view, yielding correct phase reconstruction results as shown in Fig. 8(g1) and 8(g2), where the structure of the sperm cells including tails and heads can be clearly observed.

Our Retinex DPC algorithm is not limited by a paricular imaging system and can be applied to all DPC applications. To illustrate this, we performed Retinex DPC in single-shot DPC experiments [11, 15, 17] as shown in Fig. 9 (a) where the RGB raw image is captured in a single-shot. Reconstruction results for TV-DPC and Iso-DPC are shown in Fig. 9(b) and 9(c). Results for TV-$L_2$-Retinex and TV-$L_1$-Retinex ($\gamma = 0.001$) are presented in Fig. 9(d) and 9(e) both showing their effect on background rectifying.

The Retinex DPC can also easily be combined with other penalty functions in a plug-and-play manner. Fig. 9(f) shows the combination of $L_1$-Retinex DPC with Dark-field sparse prior. With a little increase in algorithm complexity, we are able to gain an even better phase reconstruction.

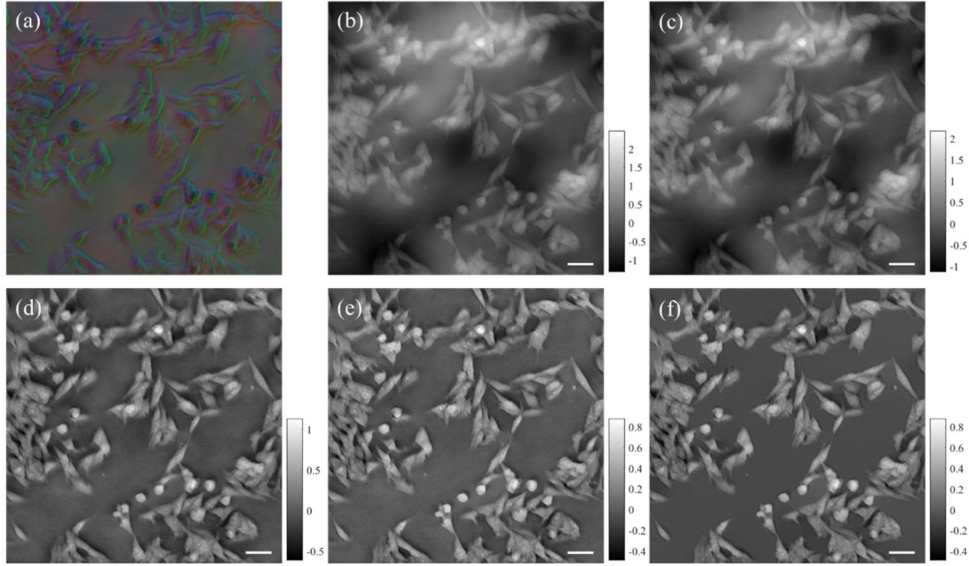

Fig. 9. Single-shot DPC experiment for lung cancer cells $NA = 0.3$, $\times 10$, pixel size is 4.2 μm. (a) Input RGB image. (b) Results for TV-DPC. (c) Results for Iso-DPC. (d) Result for TV-$L_2$-Retinex DPC. (e) Result for TV-$L_1$-Retinex $\gamma_0 = 0.05$. (f) $L_1$-Retinex + Dark-field sparse prior DPC, $\gamma_0 = 0.02$, $\beta = \alpha / 20$.

## 5. Discussion

In this research, we aimed to solve the background mismatch problem in DPC experiment and proposed Retinex DPC in which the data fidelity term in the traditional DPC reconstruction algorithm is replaced by the $L_2$-norm or $L_1$-norm of the difference between the gradient of the images and the gradient of the forward model, yielding $L_2$-Retinex DPC in Eq. (11), and $L_1$-Retinex DPC in Eq. (12). Since the Retinex DPC utilizes sharp edge information for the data fidelity, it is able to bypass the impact of background

By minimizing the cost function with Retinex data fidelity, both methods can suppress the background signal and further improve the phase reconstruction quality, while $L_1$-Retinex DPC is more difficult to be solved than the $L_2$-Retinex DPC, but it obtains stronger background suppression than that of $L_2$-Retinex DPC due to the sparse promotion of $L_1$-norm. According to our experiment, we would recommend choosing $L_2$-Retinex DPC for rapid and easy-use DPC deployment when the background problem is not prominent while using $L_1$-Retinex DPC when the background problem is severe for high-fidelity phase reconstruction.

If only Tikhonov-penalty is used, the $L_2$-Retinex DPC can be solved through two times of Fourier transform, while the $L_1$-Retinex DPC still needs to be solved iteratively. When the TV-penalty is involved, both $L_2$- and $L_1$-Retinex DPC need to be solved iteratively, but the computational complexity is actually similar to Eqs. (15), (16), and (21) can be parallel calculated.

The $L_1$-Retinex DPC uses the split Bregman method, while actually introducing two additional parameters $\gamma$ for the original $L_1$-norm term and $\gamma_0$ for the quadratic penalty. Fortunately, the split Bregman method is a mature technique in image processing, and it has been proved that $\gamma_0 = 1$ can yield correct and precise/stable convergence of the iteration [26]. In general, the $L_1$-Retinex DPC introduces only one parameter $\gamma$ that needs to be manually adjusted.

Although we only used the TV-penalty in our research, the Retinex DPC changes the data fidelity term in Eq. (1) which makes it compatible with other penalty functions as listed in Tab. 1, as the penalty function describes the phase's own statistical property. For problematic background where it changes very fast, the Retinex DPC may unable to suppress the background, in these cases, we may choose different penalty functions such as higher-order TV-penalty or the dark-field sparse prior.

## 6. Concluding remarks

We introduced Retinex theory into DPC reconstruction, and propose two new cost function for DPC deconvolution, which can suppress the impact of background mismatch that can usually happen in DPC experiments. Our model is based on replacing the traditional data fidelity term by the gradient of images and forward model, which enables background rectified DPC deconvolution, while maintains the reconstruction fidelity. Examples and statistics demonstrate the performance of our method.

Our Retinex DPC will decrease the complexity of DPC experiment as it does not requires any modification on optical systems, and further improve the experimental robustness. No background calibration is needed, and a good phase pattern can be reconstructed even the raw data is severely degenerated by the Background problem. This algorithm can be applied to all DPC experiments including but not limited to single shot DPC [15, 17, 27-29], 3D-DPC [30], and related applications [31].

**Declaration of Competing Interest**: The authors declare that they have no known competing financial interests or personal relationships that could have appeared to influence the work reported in this paper.

## Acknowledgments

This research was funded by Universiteitsfonds Limburg SWOL 2022 (CoBes22.041); Natural Science Foundation of Anhui Province (1908085MA14); Research Fund of Anhui Institute of Translational Medicine (2021zhyx-B16); Key Research and Development Program of Anhui Province (2022a05020028).